\documentclass[onecolumn,showkeys,reprint,superscriptaddress,nofootinbib,amsmath,amssymb,aps,floatfix,]{revtex4-1}
\usepackage{amsmath,amsfonts,amssymb}
\usepackage{subfigure}
\usepackage{graphics,graphicx}
\usepackage{xcolor}
\usepackage{hyperref}
\usepackage[a4paper, total={7in, 10in}]{geometry}

\begin{document}

\title{Temporal properties of higher-order interactions in social networks}

\author{Giulia Cencetti}
 \affiliation{Mobs Lab, Fondazione Bruno Kessler, Via Sommarive 18, 38123, Trento, Italy}

\author{Federico Battiston}%
\affiliation{Department of Network and Data Science, Central European University, 1100, Vienna, Austria}

\author{Bruno Lepri}%
 \affiliation{Mobs Lab, Fondazione Bruno Kessler, Via Sommarive 18, 38123, Trento, Italy}%
 
\author{M\'{a}rton Karsai}
\email{karsaim@ceu.edu}
\affiliation{Department of Network and Data Science, Central European University, 1100, Vienna, Austria}

\begin{abstract}
Human social interactions in local settings can be experimentally detected by recording the physical proximity and orientation of people. Such interactions, approximating face-to-face communications, can be effectively represented as time varying social networks with links being unceasingly created and destroyed over time. Traditional analyses of temporal networks have addressed mostly pairwise interactions, where links describe dyadic connections among individuals. However, many network dynamics are hardly ascribable to pairwise settings but often comprise larger groups, which are better described by higher-order interactions. Here we investigate the higher-order organizations of temporal social networks by analyzing three publicly available datasets collected in different social settings. We find that higher-order interactions are ubiquitous and, similarly to their pairwise counterparts, characterized by heterogeneous dynamics, with bursty trains of rapidly recurring higher-order events separated by long periods of inactivity. We investigate the evolution and formation of groups by looking at the transition rates between different higher-order structures. We find that in more spontaneous social settings, group are characterized by slower formation and disaggregation, while in work settings these phenomena are more abrupt, possibly reflecting pre-organized social dynamics. Finally, we observe temporal reinforcement suggesting that the longer a group stays together the higher the probability that the same interaction pattern persist in the future. Our findings suggest the importance of considering the higher-order structure of social interactions when investigating human temporal dynamics.
\end{abstract}

\maketitle

\section*{Introduction}

\begin{figure*}
\centering
\includegraphics[width=0.7\textwidth]{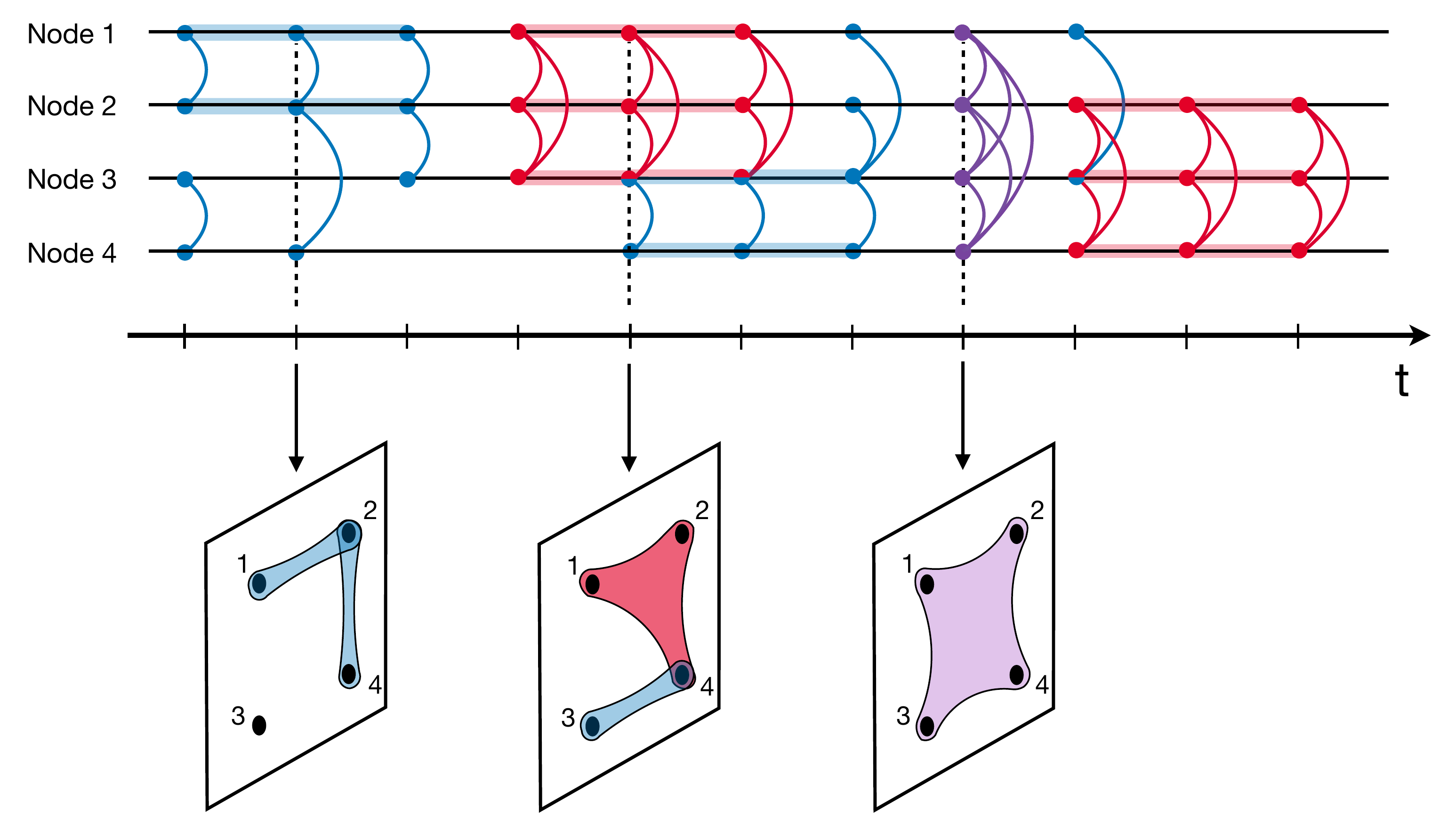}
 \caption{\textbf{From higher-order interactions to temporal hypergraph.} Example of higher-order interactions among a group of four people. The horizontal lines represent the temporal behavior of each individual and  curved lines bridge the $n$ nodes involved in one interaction, with different colors for different sizes: blue for interaction between two people (2-hyperedges), red for three people (3-hyperedges) and purple for four people (4-hyperedges). Coloured lines on individual timelines indicate the time and duration of interactions, with colors coding their size. Snapshots indicate corresponding hypergraphs at specific times. Note that as the data describe face-to-face interactions and not co-location, it is possible to observe open structures, like for instance open triads, i.e. one person interacting with two people that are not connected to each other.
 }
 \label{fig_panel_hyper}
\end{figure*}


Complex networks are fundamental tools to represent complex systems made of interacting units, with applications in biology, social sciences, transport infrastructures, communications, financial markets, and more~\cite{newman2018networks, barabasi2016network,latora2017complex}.
Incorporating a set of discrete nodes and the connections  between them, the networks schematize the existing relationships among agents, providing a synthetic picture of the system architecture. Despite the success of network representations of complex systems in the last thirty years, static graphs fall short to effectively describe a wide variety of real world systems, especially when the dynamics of their structural changes is in focus.
In networked systems, whether nodes represent people, cells, neurons, virtual or physical sites, their interactions are not bounded to be static but are rather evolving, with nodes and links, which appear and disappear over time.
To address the time-varying aspect of complex structures, the field of temporal networks emerged~\cite{holme2012temporal,lambiotte2016guide,holme2015modern,latapy2018stream} providing useful representations and tools to study the dynamics of real complex systems. The framework is particularly suited to describe social systems where coupling contacts among people naturally change over time in online and offline social networks, email and mobile phone communications, and more~\cite{karsai2011small,miritello2011,zhao2011,karsai2014,kobayashi2019}. However, social interactions may vary over multiple temporal scales, ranging from  long lasting friendships to accidental interactions between strangers. Moreover, consecutive interactions may not appear independently but follow each other rapidly forming bursty patterns~\cite{karsai2018bursty} potentially due to intrinsic correlations~\cite{barabasi2005origin} or simply via circadian fluctuations of human activity~\cite{malmgren2008poissonian}. Temporal networks describe such processes at the highest time resolution to understand how single interactions may lead to collective phenomena, as long trains of bursty events, or the emergence of the complex social structure.

Network approaches were originally devised to describe dyadic relationships and can only provide a limited representation of systems interacting beyond pairwise schemes. Such higher-order interactions are ubiquitous~\cite{battiston2020networks}, from human societies to artificial or biological systems. For instance scientific authors naturally team up in larger groups to complement the expertise of different members~\cite{patania2017shape}, neurons send and receive stimuli from multiple adjacent partners at the same time~\cite{petri2014homological, giusti2016two}, and the stability of large ecosystems relies on mutual and cooperative partnerships often involving three or more species~\cite{grilli2017higher, bairey2016high}.
Besides, higher-order interactions were shown to significantly modify the collective behavior of many dynamical processes, from diffusion~\cite{schaub2020random,carletti2020random} and synchronization~\cite{skardal2019abrupt,millan2020explosive,lucas2020multi} to spreading~\cite{iacopini2019simplicial, de2020social}, social dynamics~\cite{neuhauser2020multibody} and games~\cite{alvarez2020evolutionary}. For a thorough introduction on the structure and dynamics of these higher-order systems, we refer the interested reader to the  comprehensive overview provided in Ref~\cite{battiston2020networks}.

In this paper our goal is to study the heterogeneous dynamics of group interactions by looking at bursty patterns of higher-order structures in temporal networks. We analyze the temporal properties of multi-party face-to-face interactions~\cite{genois2018can} recorded in the SocioPatterns project~\cite{isella2011}. We define group interactions in this setting and determine the number of groups to classify them according to their size. By analysing their duration and the time between their subsequent appearances we identify long bursty trains of recurrent group interactions due to temporal correlations.
Finally, we investigate the temporal evolution of groups and how their size changes over time by progressively acquiring or losing members, observing a reinforcement of group structures over time. Our results generalise universal phenomena earlier observed for dyadic interactions~\cite{karsai2012universal} for the case of higher-order temporal structures.




\section*{Results}

\begin{figure*}
\centering
\hspace*{-9mm}
\includegraphics[width=0.9\textwidth]{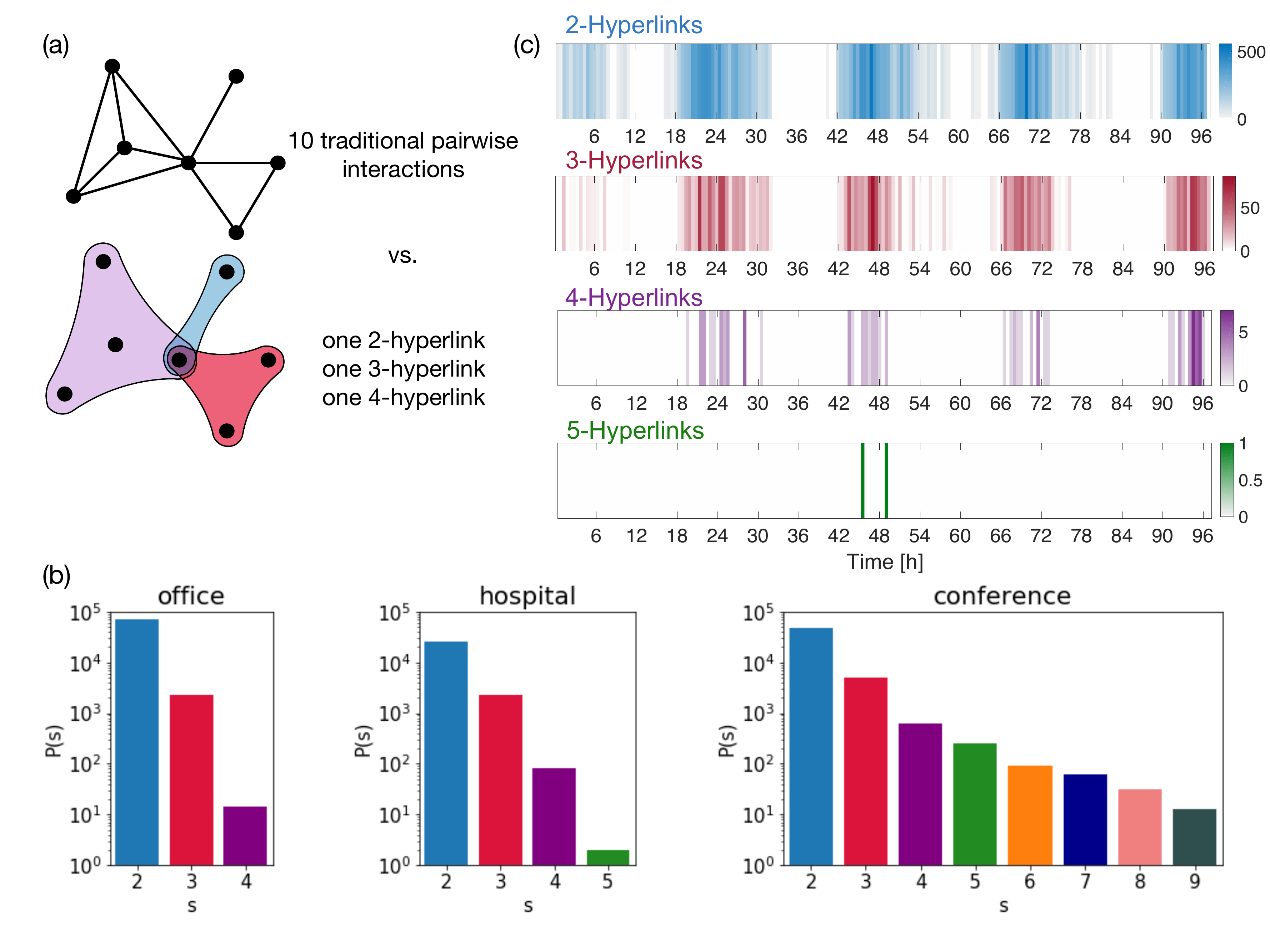}
 \caption{\textbf{Higher-order structure of temporal human interactions.} (a): example of network depicted in a traditional way or as a hypergraph. In the last case interactions of different sizes correspond to different colors: blue for 2-hyperedges (two nodes), red for 3-hyperedges (three nodes), purple for 4-hyperedges (four nodes). (b): histograms reporting the counts of interactions for each different size in the three datasets of face-to-face human communication: an office, an hospital and a scientific conference. (c): time series of interactions in the hospital dataset represented by hyperedges of size2, 3, 4, and 5. }
 \label{fig_panel1}
\end{figure*}

\subsection*{Temporal higher-order social interactions}
We aim at investigating the temporal dynamics of the higher-order structure of human proximity interactions in different social settings. To this end, we choose three datasets, which describe face-to-face interactions~\cite{isella2011,genois2018can} (a) in an office building in France~\cite{genois2015} over 11 days; (b) in a hospital ward between patients, medical doctors, nurses and administrative staff over 72 hours~\cite{vanhems2013estimating}; and (c) during 32 hours in a scientific conference~\cite{isella2011}. Each dataset records the dyadic face-to-face interactions of people with time resolution of $20$ seconds, but also identifies simultaneous contacts of the participants thus allowing for the observation of group interactions. Originally exploited for pairwise network analysis, the fine grained temporal structure of these interactions allows us to reconstruct the formation, presence, and deletion of higher-order groups. 


In the traditional network formalism, a dyadic temporal interaction between two people $a$ and $b$ at time $t$, which lasted for  duration $d$, is represented by a temporal link $e=(a,b,t,d)$. In this setting, the sequence of temporal events builds up a temporal network $G_T=(V_T,E_T,T)$, where any node $a\in V_T$, any event $e\in E_T$ and the network evolve over $T$ period, thus $0\leq t \leq T$ and $0\leq d \leq T$.

However, people often connect in larger groups, where more than two individuals interact at the same time. Simple links, by definition describing dyadic connections, are not suited to describe such higher-order interactions, which require different types of building blocks, known as {\it hyperedges}. An $n$-hyperedge, or hyperedge of size $n$, describes an interaction of $n$ individuals. In more mathematical terms this is denoted a simplex of order $n-1$~\cite{battiston2020networks}.
Simple dyadic links represent the first non-trivial interaction, described by a 2-hyperedge. For temporal data, we define the interaction between a group of $n$ people, $i_1, \ldots, i_n$, at time $t$ and for duration $d$ as a temporal $n$-hyperedge assigned as $e_{n}=(i_1, \ldots, i_n,t,d)$. The sequence of temporal events builds up a temporal hypergraph $H_T=(V_T,E_T,T)$, where any node $i\in V_T$, any event $e_n\in E_T$ (now describing a set of hyperedges) and the hypergraph evolve over $T$ period, thus $0\leq t \leq T$ and $0\leq d \leq T$. An example of a temporal hypergraph is shown in Fig.~\ref{fig_panel_hyper}, where the connections that nodes undertake are coloured according to their size shown with some instantaneous snapshots of the temporal hypergraph underneath.

In the considered datasets each interaction is originally stored through simple links. However, they do not necessarily represent the original building blocks of the interactions. The fine-grained temporal nature of the datasets allow us to reconstruct the original higher-order features of the connections and the corresponding hyperedges. In practice, if at a time $t$ there are $n*(n+1)/2$ dyads between the members of a set of $n$ nodes such that they form a fully connected clique, we promote the $n*(n+1)/2$ links to a $n$-hyperedge. For instance, if at time $t$, $a$ is interacting with $b$ and $c$, and $b$ is interacting with $c$ too, the interactions will be stored into an 3-hyperedge. Note that the same reconstruction is not possible from temporally-aggregated data, where the presence of a closed triangle may be the byproduct of the temporal aggregation of distinct truly pairwise interactions. This is demonstrated in Fig.~\ref{fig_panel1}(a) where the schematic representation of a network is compared with its hypergraph version. The traditional network is characterized by ten simple links, while considering their simultaneous group interactions, we identify one 2-hyperedge (in blue), one 3-hyperedge (in red), and one 4-hyperedge (in purple).

In the following when we refer to a group interaction of size $n$, it corresponds to a maximal clique of size $n$, or in other words an $n$-hyperedge, without being a part of any larger group.



\subsection*{Statistics of higher-order interactions}

The statistics of \textit{maximal} higher-order interactions for the three datasets are reported in Fig.~\ref{fig_panel1}(b). Smaller interactions involving less people are more numerous in all datasets, however different settings are characterized by different statistics: for instance the conference dataset reveals the presence of very large aggregations, with up to events of size 9, while in the hospital and office settings the group sizes are limited to 5 and 4 respectively. Note also that the office dataset was collected for a longer period thus it represents the most connected aggregated network and the one with the largest total number of interactions. However, these interactions are mainly pairwise, as demonstrated in Fig.~\ref{fig_panel1}, which is particularly peaked at $s=2$, while interactions of size 4 are poorly represented (less than 20 in 11 days). In addition, the office network has the highest ratio (nearly one and half orders of magnitude) between the number of dyadic and triplet interactions. This suggests that the office network is ``the lowest-order", especially as compared to the conference network, which instead appears to be the ``the highest-order" network. In general, the presence of several group interactions in these networks and their heterogeneous size call for a deeper analysis of their higher-order structures to properly characterize their dynamical evolution.

Moreover, it is interesting to observe that the emergence of higher-order structures is strongly heterogeneous in time. This is evident from Fig.~\ref{fig_panel1}(c) where we show the timely occurrences of interactions of sizes 2, 3, 4 and 5 in the hospital dataset. Note that similar time-series for the other datasets are reported in the Supplementary Information. This visualisation already suggests bursty patterns of higher-order interactions, which are not independent across different orders. In one way it is not surprising as higher-order events always build up from lower-order structures, but their heterogeneous dynamics and short term recurrence is far from being obvious. In the following section we will provide a more formal inspection of these features by defining and analysing higher-order bursty behavior.


\subsection*{Higher-order bursty behavior}

\begin{figure*}[ht!]
\centering
\includegraphics[width=0.9\textwidth]{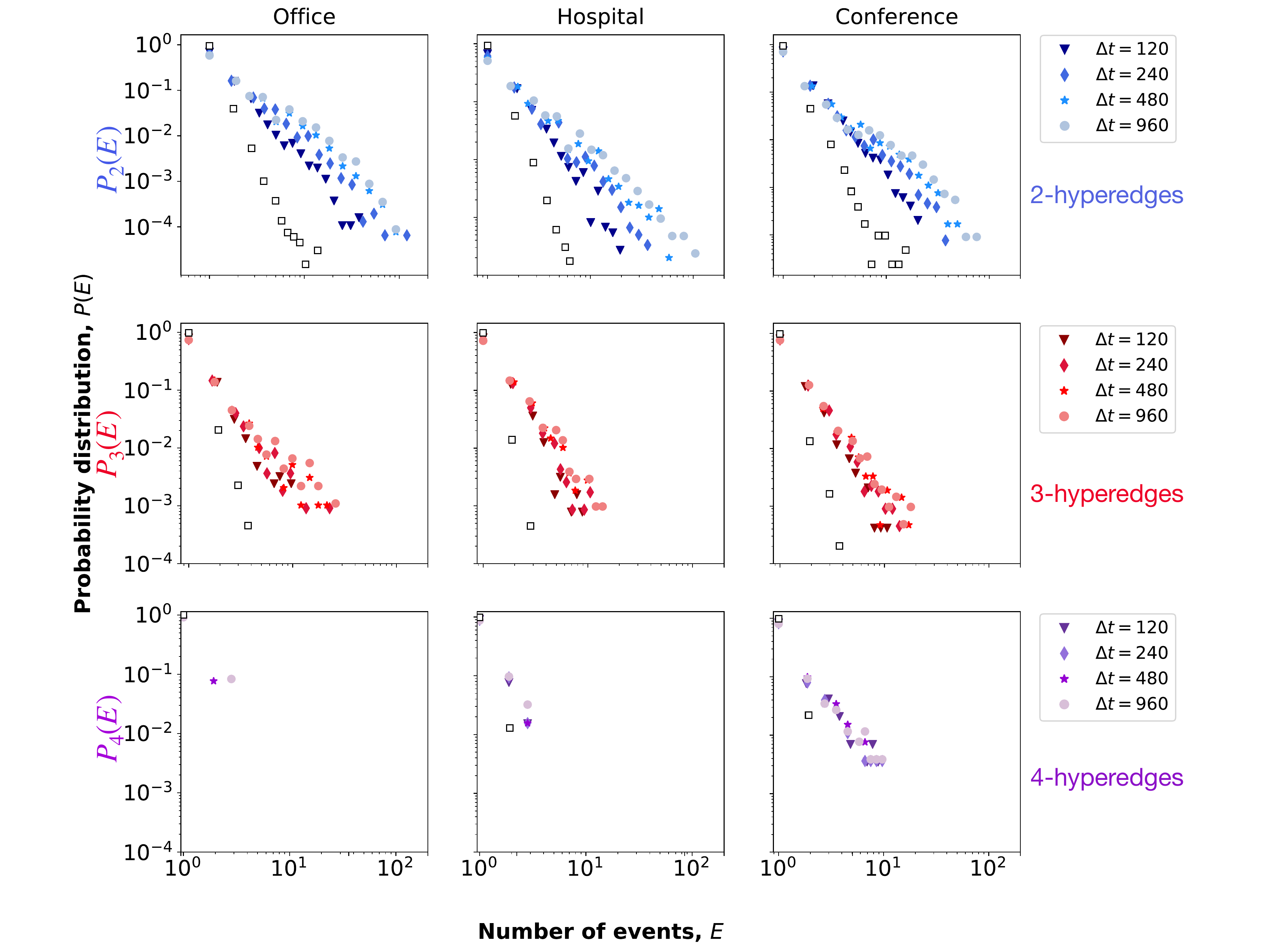}
 \caption{\textbf{Higher-order structures of temporal trains.} Number of events distribution $P(E)$ for group interactions of fixed size: size 2 (first line), size 3 (second line) and size 4 (third line) interactions. Symbols and colors represent different values of the aggregation window $\Delta t$. Empirical results (full symbols) are compared to the null models obtained by shuffling the event times (empty squares, distributions obtained aggregating with $\Delta t=120$ seconds).}
 \label{PE}
\end{figure*}

To study the dynamics of higher-order interactions we study the dynamics of events, which can be a singular interaction or hyperedge of any kind and duration.
In the investigated datasets interactions were recorded every 20 seconds, which define the minimum duration associated to an interaction. In order to construct events with longer duration, we merge consecutive events which involved exactly the same group of people. In this way we are able to identify longer events with durations modulo 20 seconds.

Another interesting quantity measures the time between consecutive events of the same group of people. More precisely, if a generic event $i$ begins at time $t_i$ and has duration $d_i$, inter event time $t_{ie}$ is defined as $t_{ie}=t_{i+1}-(t_i+d_i)$. In other words it spans from the end of the group's previous interaction to the beginning of the next one. Inter-event times are a central measure to study event dynamics as their distribution evidently show whether the dynamics are heterogeneous and thus indicated by a broad $P(t_{ie})$, or they resemble a homogeneous dynamics, such as a Poisson process, with exponential inter-event time distribution~\cite{karsai2018bursty}. In the Supplementary Information, we report the probability density functions of event durations and inter-event times, respectively in Figs.~4 and 5. These results show evidently that face-to-face interactions are strongly heterogeneous in duration and inter-events times regardless the social setting.

\begin{table}[]
    \centering
    \begin{tabular}{c|c|c|c}
         & Office & Hospital & Conference  \\
        \hline
         size 2 & 0.58 & 0.61 & 0.58\\
         size 3 & 0.63 & 0.54 & 0.62\\
         size 4 & -0.17 & 0.79 &  0.70\\
    \end{tabular}
    \caption{Burstiness measure for distributions $P_n(E)$ reported in Fig.~\ref{PE}. The burstiness has been computed according to the formula firstly proposed in \cite{goh2008burstiness} and successively normalised~\cite{kim2016measuring} in order to allow a comparison between samples with different number of events.}
    \label{table_B}
\end{table}

To further quantify burstiness in event sequences of different size we measured the burstiness index, defined in Ref.~\cite{kim2016measuring} as $B = [\sqrt{n+1}r - \sqrt{n-1}]/[(\sqrt{n+1} - 2)r + \sqrt{n-1}]$, where $r = \sigma_{\tau}/\langle \tau \rangle$ with $\langle \tau \rangle$ the mean inter-event time and $\sigma_{\tau}$ the corresponding standard deviation respectively. This measure is corrected for the sample size $n$ and represents an improved version of the original measure defined in Ref.~\cite{goh2008burstiness}. This index takes values between $B=-1$ for regular signals, $B=0$ in case of independent events, and $B>1$ in case events are temporally correlated. Average values of $B$ for the three considered datasets and up to interactions of size 4 are reported in Table~\ref{table_B}. With the exception of  interactions of size 4 in the office setting, for which we lack sufficient statistics, all other cases presented appeared with values of burstiness significantly larger than 0. Interestingly, burstiness of events of different sizes appear to be comparable.

\begin{figure*}[ht!]
\centering
\includegraphics[width=0.9\textwidth]{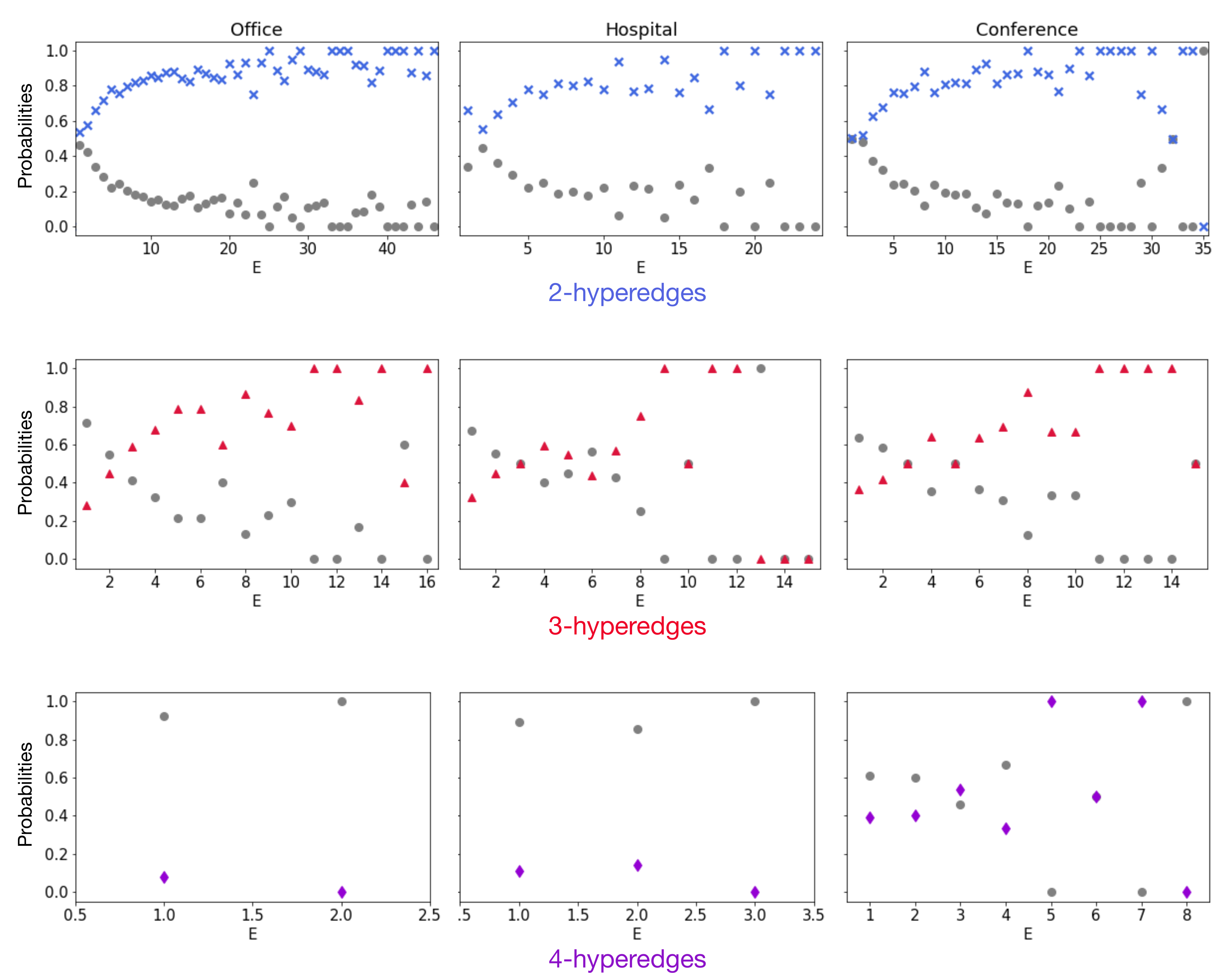}
 \caption{\textbf{Group evolution and temporal reinforcement of higher-order human interactions.} Each panel depicts the probabilities that after a train of interaction at least with $E$ events the people involved either interact again in the same or a higher-order structure (coloured symbols) or they do not reconnect anymore (grey symbols) within $\Delta t$. Panels (a)-(c) show results for dyadic trains; (d)-(f) for triadic (3-hyperedge) trains; and (g)-(i) for 4-body (4-hyperedge). Results are shown for the three analysed datasets for trains identified with $\Delta t = 480$ second.}
 \label{evol}
\end{figure*}

The time series reported in Fig.~\ref{fig_panel1} anecdotally suggest that events often occur in successions of high activity, known as \textit{trains of events}, alternated with periods of inactivity. This phenomenon has already been observed for pairwise interactions in various temporal processes~\cite{karsai2012universal}, like communication (i.e. emails, text messages or mobile calls), recurrent seismic activities in a specific location, and neuron firing signals. It has been argued in Ref.~\cite{karsai2012universal} that the emergence of long bursty trains is ascribable to short-term temporal correlation between events. This can be demonstrated by the distribution of the $E$ number of events in the bursty period. To define $E$ we need to identify events, which belong to the same bursty period, also called bursty train. In our definition we consider two events to be related if they are consecutive and happen with an inter-event time smaller than a given value $\Delta t$. Related consecutive event pairs can build up to longer trains where the above condition is true for each consecutive event and otherwise the train is separated by longer than $\Delta t$ inter-event times from the rest of the sequence. The number of events in these trains give the metric $E$, which distribution appears as exponential in case of independent events, while any deviation from this scaling indicates present temporal correlations between the events in the trains. In empirical observations, as mentioned before, the $P(E)$ distribution has been found to be well approximated by power-law functions, evidently indicating temporal correlations characterising these systems~\cite{karsai2012universal}.

However, bursty event trains have never been investigated for events involving more than two nodes. Here we move beyond traditional pairwise interactions and we separate the events according to their size to identify trains of events of each order separately. As earlier defined, we introduce a parameter, $\Delta t$, which allows to discern what we consider related events from uncorrelated ones and to identify event trains. We can identify trains containing only events of a specific size $n$ and compute their quantity $E$ to obtain the distribution $P_n(E)$ for events of size $n$. Such distributions computed for different event sizes and datasets appear with heavy tails, as shown in Fig.~\ref{PE}. Moreover, this phenomena appears to be robust against the choice of $\Delta t$ values, coherently with the analysis presented in Ref.~\cite{karsai2012universal}. Note that these observations cannot be reproduced by simple null models where event sequences are constructed from uncorrelated interactions obtained by shuffling event times. Equivalent distributions computed in such independent signals are shown on panels of Fig.~\ref{PE} as empty symbols, appearing evidently different than the empirical observations. For further details on the definition of utilised null models see Methods and Ref.~\cite{karsai2012universal}

In summary, these results indicate the existence of bursty dynamics not only for dyadic but also for higher-order event sequences. We observed that they evolve in bursty trains of correlated events in case of any size and investigated dataset. More importantly these observations cannot be reproduced by null models of independent events, indicating the observed correlations to be significant in the empirical systems.

\subsection*{Evolution and formation of higher-order social interactions}


\begin{figure*}[ht!]
\includegraphics[width=0.7\textwidth]{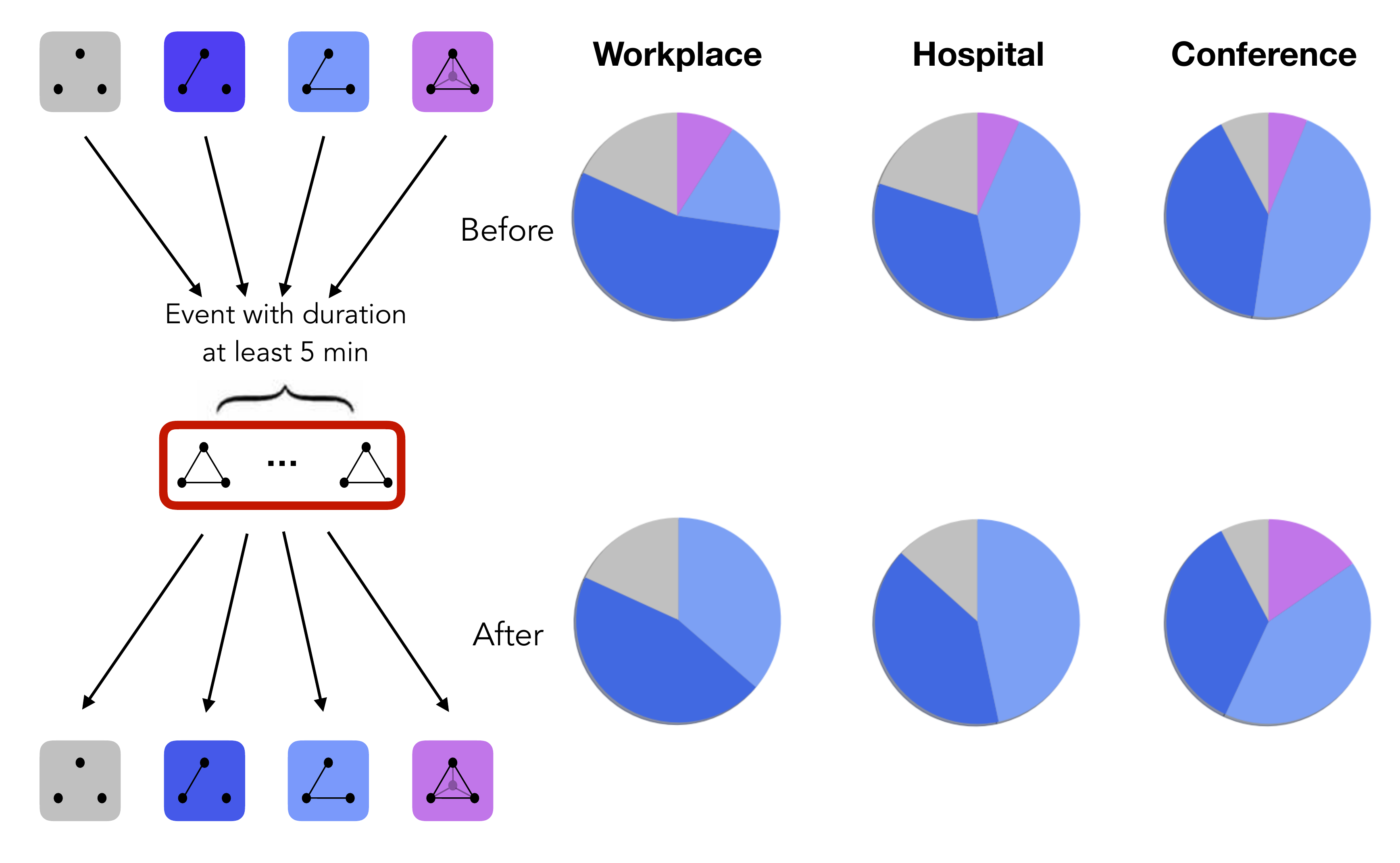}
\caption{\textbf{Transition rates of higher-order configurations.} Configurations before and after a triplet event lasting at least 300 seconds. Four classes of configurations are possible for three nodes that are not part of a 3-hyperedge: all disjoint (grey configuration), one or two pairwise interactions (blue configurations) or they are all connected and part of larger interaction (purple configuration).}
 \label{fig_bef_aft}
\end{figure*}

The above  analyses allowed to generalize to higher-order network measures some findings firstly observed in dyadic settings. However, this framework also allows for some genuine higher-order investigation about group formation and evolution, similar to Ref.~\cite{benson2018simplicial}. The composition of a group in general is related to the previous interaction history of participants, which extends well beyond pairwise relationships. To observe any group formation scenario, we look for the presence of actual interactions in forthcoming time steps thus following how the order of events changes in time. In practice, for events of a given size $s$ we consider all the trains of size $E$, and measure the probability that (a) the train continues with an $E+1$th event of the same or higher-order or (b) the group falls apart. The results are shown in Fig.~\ref{evol} for our three datasets, extending to higher-order interactions a similar analysis proposed in Ref.~\cite{karsai2012universal} for traditional pairwise communications. Panels in the first line of Fig.~\ref{evol} depict the evolution of dyadic interactions. For each value of $E$ blue crosses indicate the probability that the event is followed by a new event of the same or higher-order, while complementary probability, shown as grey dots, measures the probability the corresponding nodes break their interaction in the following time step. Analogous measures are shown in the second and third line for interactions of size 3 and 4, where the two probabilities are shown in color and in grey respectively. Overall, the colored symbols display an increasing trend across the different interaction sizes and datasets. These results indicate the existence of temporal reinforcement, meaning that the longer the length of an interaction -- no matter the group size -- the higher the chances the relationship will not break down. We note that these trends are more pronounced for groups of small size, which could be due to the significantly larger number of smaller size events in the networks. Although these observations are rooted in some earlier results on group formation dynamics observed through dyadic interactions~\cite{zhao2011social,karsai2012universal,karsai2012correlated}, they provide an independent verification of similar phenomena by using higher-order events.


The question remains, what happens exactly before and after a higher-order event  is formed? To answer this question we depart from the exclusive investigation of higher-order events. 
For each event corresponding to a hyperedge of generic size $n$, we identify the relations among  the $n$  nodes involved one step before the formation of the group and one step after it disappeared. Out of simplicity, we focus on groups of size 3 (3-hyperedges), as they are by far more numerous within our datasets. At the previous and following time step a clique of three nodes can be arranged across for different classes, as illustrated in  Fig.~\ref{fig_bef_aft}. In the first case there are no connections between any of the nodes (grey sketch); in the second one there is a single link connecting two of the three nodes, while the last unit is disconnected (dark blue); in the third case the nodes are connected across an open triad (light blue); the fourth configuration (purple) represents the case where the three nodes are interacting all together but they are part of a larger hyperedge (and for this reason they are not classified anymore as a maximal 3-hyperedge). 
The pie charts in Fig.~\ref{fig_bef_aft} depict the proportions of the four different configurations before and after a higher-order interaction of size 3 for events lasting at least 300 seconds.

Results in Fig.~\ref{fig_bef_aft} suggest that a triplet is infrequently formed from or evolve into a larger group, as purple sections indicate scarce observation of this case in Fig.~\ref{fig_bef_aft}. Exception is the conference setting, where events growing to higher-order seem to be more frequent. At the same time, it is rather infrequent, especially in the conference setting, that a group is created from scratch or vanishes into three isolated nodes (grey configuration are also quite uncommon). The configurations that prevail are those where two nodes are already connected and a third one is added (dark blue) or, alternatively, two nodes are linked to the same node in an open triad and then they get connected (light blue). Similarly, transition rates for the dis-aggregation of the interaction are high when the triplet is broken in one (dark blue) or two (light blue) connected couples. 

These results suggest a more similar group formation dynamics in the interaction pattern of the hospital and the office settings, where people may undergo work dynamics, dictated by daily work routines. Observation may be driven by scheduled meetings where a group of people come together suddenly at a given time and then depart. Group formation, instead, appears to be more fluid at the conference, where individuals can connect more freely thus it is more common for groups to aggregate and disaggregate step by step, one node at a time. Another important difference between the two kinds of datasets is the formation frequency of larger cliques, indicated by the purple section in Fig.~\ref{fig_bef_aft}. Indeed, in the first two datasets (i.e. the hospital and the office) a triplet can in few cases stem from the disintegration of larger groups, but the opposite, i.e. a triplet increasing its size by acquiring new members, never happens. This last possibility is instead common in the conference dataset, where the probability that a group size switches from 3 to 4, or even more, is even higher than the opposite, i.e. a triplet generated from  a larger group. This suggests that a triplet is more suitable to represent the starting point of a larger aggregation in the conference setting than in the two working places, and therefore the greater tendency in the former environment to build groups step by step.



\section*{Conclusions}

In this work we investigated the dynamics of higher-order interactions in temporal social networks. To this scope, we made use of three publicly available datasets of face-to-face human interactions collected in different settings as in a hospital, in an office, and during a conference. Originally analyzed by means of traditional network tools, the temporal nature of the datasets allowed us to reconstruct the real higher-order organization of social interactions. A first analysis of the datasets revealed the presence of frequent higher-order interactions not limited to simple dyads. More interestingly, such higher-order events appear with heterogeneous bursty dynamics, however with lower frequency for higher-order.

By following the time evolution of the different kinds of interactions we observed bursty trains of higher-order events in all settings. The distributions of bursty train sizes revealed a broad tailed scaling, hinting at similar behavior of higher-order interactions already observed for dyadic events in other bursty systems in biological, geological and social domains~\cite{karsai2012universal}.We also inspected memory effects in group formation by measuring the probability that a specific train of interactions is protracted in time, based on the number of previous events and its groups size. We discovered that interactions lasting longer times are more likely to persist even longer, potentially due to temporal reinforcement mechanisms characterising all settings.

Group evolution showed differences across the considered datasets. In particular for higher-order interactions involving three individuals, we looked at the relational structures at the preceding and forthcoming periods. We found in the hospital and office settings similar behavior, possibly due to their work related organisation where individuals are subject to pre-organized and regular dynamics, leading to a higher probability to generate or dis-aggregate groups instantaneously. 
Differently, in the conference setting we observed a tendency to build groups by progressively adding members, one step at a time, reflecting a more spontaneous way of group formation.

This study however comes with some limitations. First, despite the large amount of face-to-face interactions captured by the three analyzed datasets, the investigation of large group behaviors is inherently limited by the lower statistics associated to higher-order events, as compared to pairwise interactions. In the future this problem could be addressed by considering datasets with a higher proportion of non-dyadic interactions. Such a setting would indeed make possible to generalize the study of transition rates in group evolution, now presented only for triplets. Second, our study focuses on face-to-face interactions, not considering alternative types of information e.g. geo-localised data and the corresponding co-mobility networks, which are naturally suited for a higher-order analysis.

Overall, our work reveals a new level of richness in temporal human dynamics, neglected in the previous literature. We showed how, taking into account the new framework of higher-order interactions~\cite{battiston2020networks}, helps us to better characterise social dynamics extracted from three different settings. Taken together we hope that our findings will pave the way to the use of higher-order network tools for investigating the dynamics of human interactions.



\section*{Methods}

\textbf{Shuffled model.} The null model of independent sequences for traditional pairwise interactions is built from the original data by shuffling the times of the events but maintaining all the original pairwise interactions. In this way, we maintain the same time stamps and each node is interacting in the same number of times. Interested readers may look at the comprehensive review on randomized reference models in temporal networks in Ref.~\cite{gauvin2018randomized}


For group interactions, one possibility is to consider the above defined time shuffled event sequence for each pairwise interaction, and then identify higher-order interactions among them. However this method breaks almost all higher-order patterns and allows the formation of very few cliques of size larger than 2. Hence, we followed another method, where we kept each higher-order event with a given size and shuffled their occurrence times between events of the same size. This shuffling ensures that each event of a given size appears the same number of times as in the original sequence but independently from each other.


\section*{Acknowledgments}
G.C. acknowledges partial support from the “European Cooperation in Science \& Technology” (COST): Action CA15109. F.B. acknowledges partial support from the ERC Synergy Grant No. 810115 (DYNASNET). M.K. acknowledges support from the DataRedux (ANR-19-CE46-0008) ANR and the SoBigData++ (871042) H2020 projects.

\addcontentsline{toc}{chapter}{Bibliography}
\bibliographystyle{ieeetr}  
\bibliography{bib_sociopattern.bib}

\clearpage

\onecolumngrid

\appendix

\section*{Supplementary Information}

We provide a set of additional figures in order to offer the elements for a more detailed comparison between an approach taking into account higher-order behaviors and a more traditional point of view treating each singular pairwise interaction separately.

\vspace{2cm}
\centering
\subsection*{Office time series}

\vspace{2cm}

\begin{figure*}[ht]
\centering
    \includegraphics[width=0.7\linewidth]{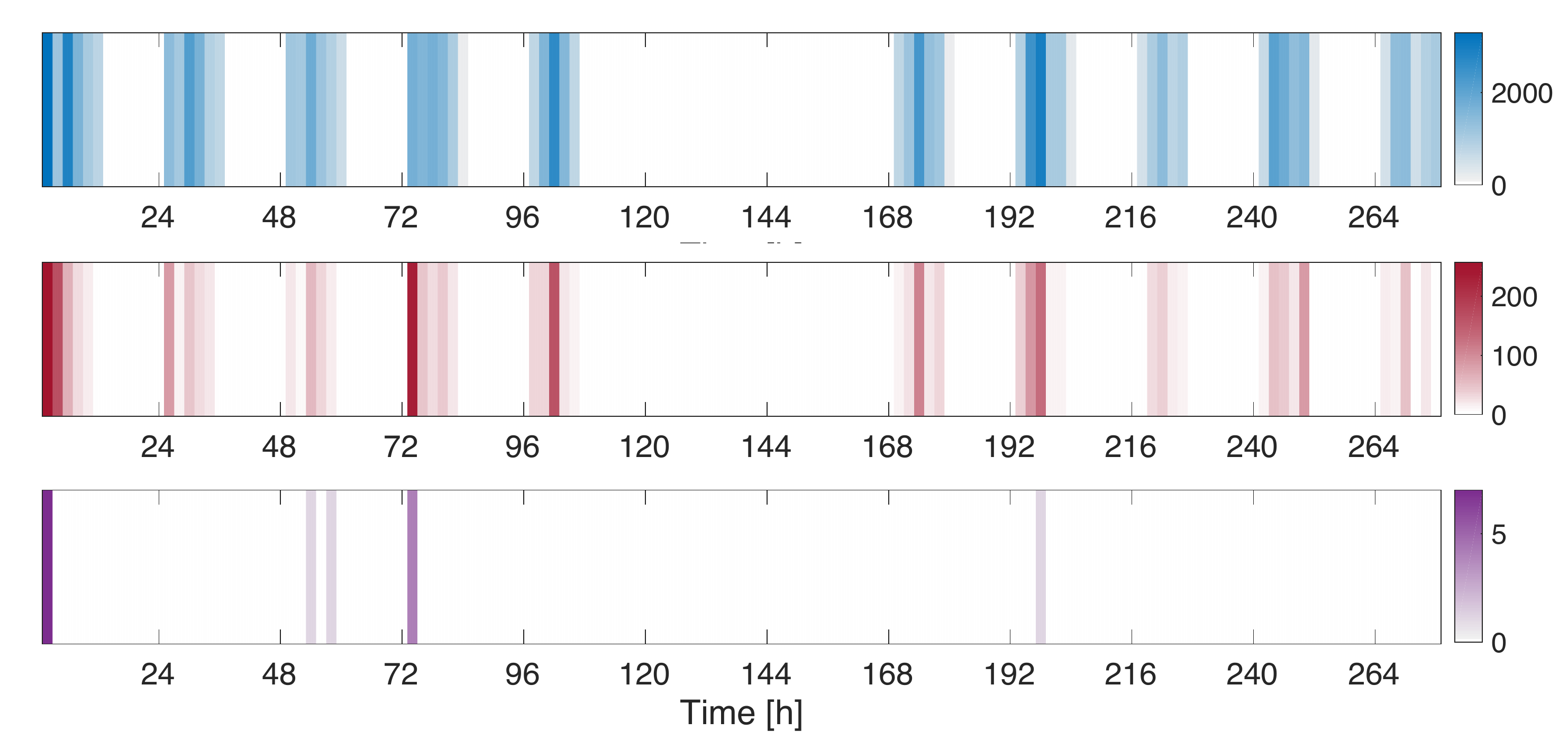}
    \caption{Time series of interactions taking place in an Office building in France~\cite{genois2015}. The data have been collected in 2015 for 11 days and the experiment has involved 217 people. The interactions are separated according to their size: blue for 2-hyperedges, red for 3-hyperedges, and purple for 4-hyperedges. We notice a 24h periodicity where couple interactions are the most frequent ones.}
\end{figure*}

\newpage

\vspace*{2cm}
\centering
\subsection*{Conference time series}

\vspace{2cm}

\begin{figure*}[ht!]
\centering
\includegraphics[width=0.7\textwidth]{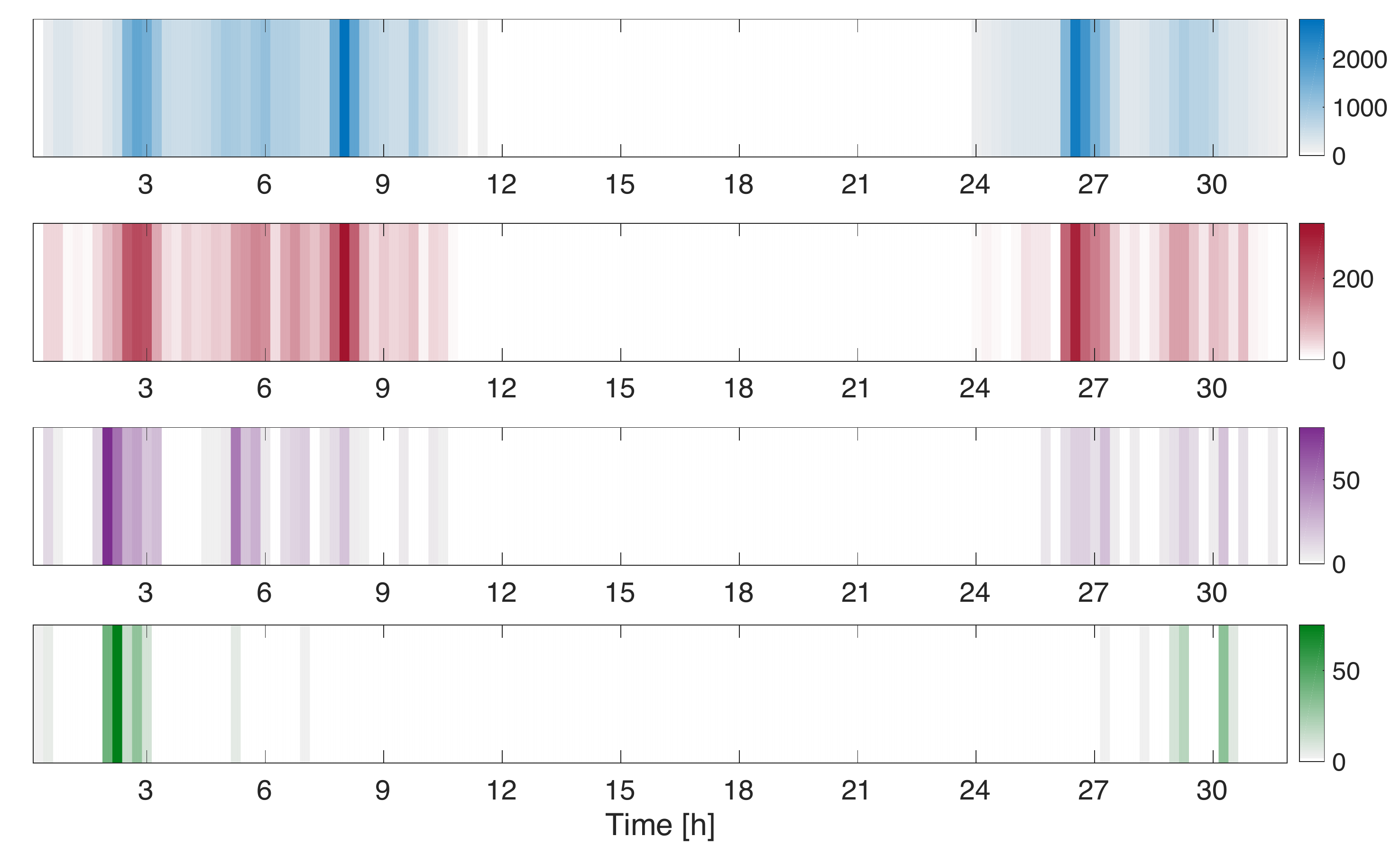}
 \caption{Time series of interactions taking place in the SFHH conference in Nice, France (June 4-5, 2009). The experiment has involved 405 people. The interactions are separated according to their size: blue for 2-hyperedges, red for 3-hyperedges, purple for 4-hyperedges, and green for 5-hyperedges. Larger group interactions are present in this dataset (up to size 9). We notice a 24h periodicity in the two days of the conference and here too we observe that smaller interactions are more frequent.}
\end{figure*}

\newpage
\vspace*{2cm}
\centering
\subsection*{Structure of temporal trains of traditional pairwise interactions}
\vspace{1cm}

\begin{figure*}[ht]
\centering
\hspace{-4mm}
\includegraphics[width=0.32\columnwidth]{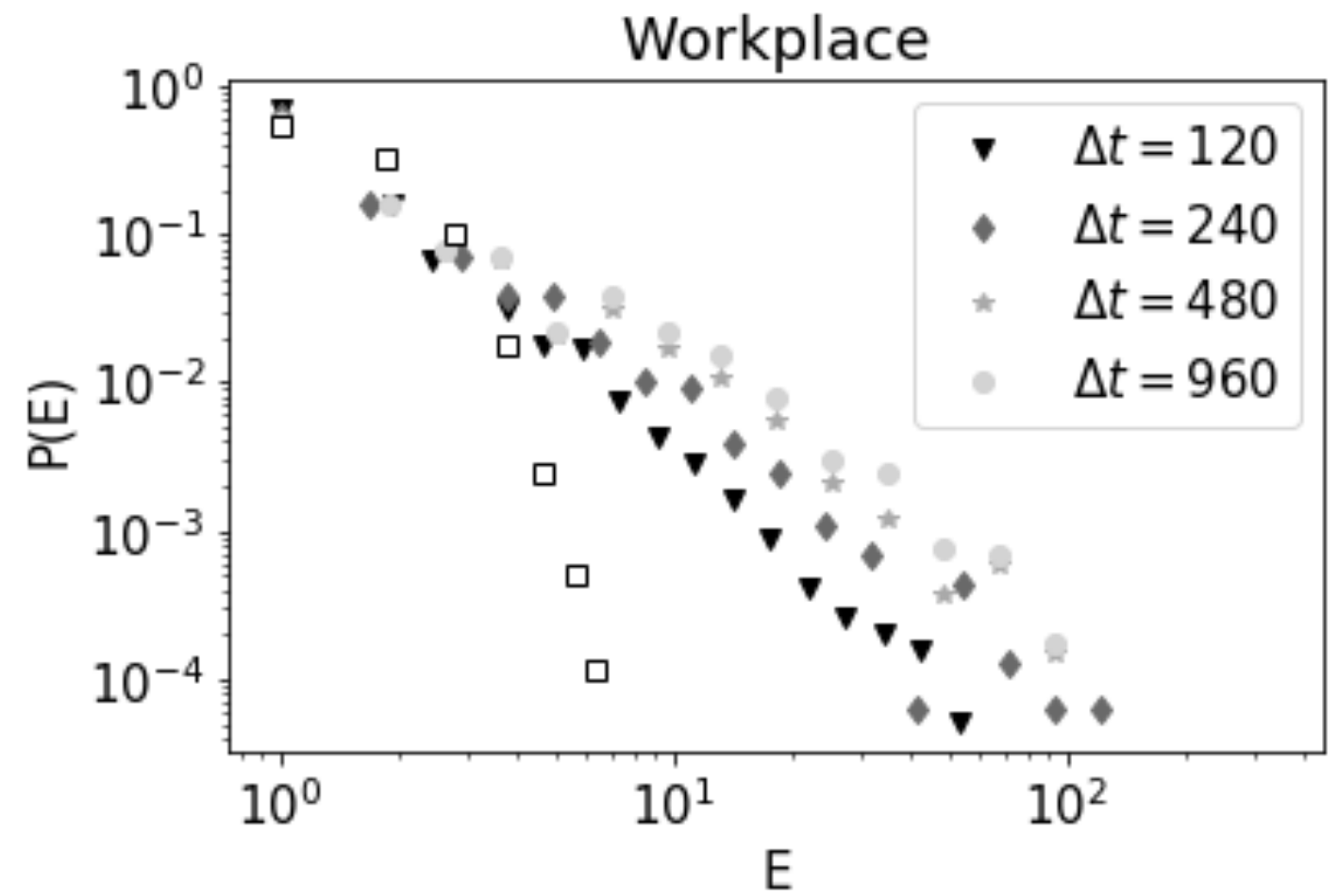}
\hspace{-1mm}
\includegraphics[width=0.32\columnwidth]{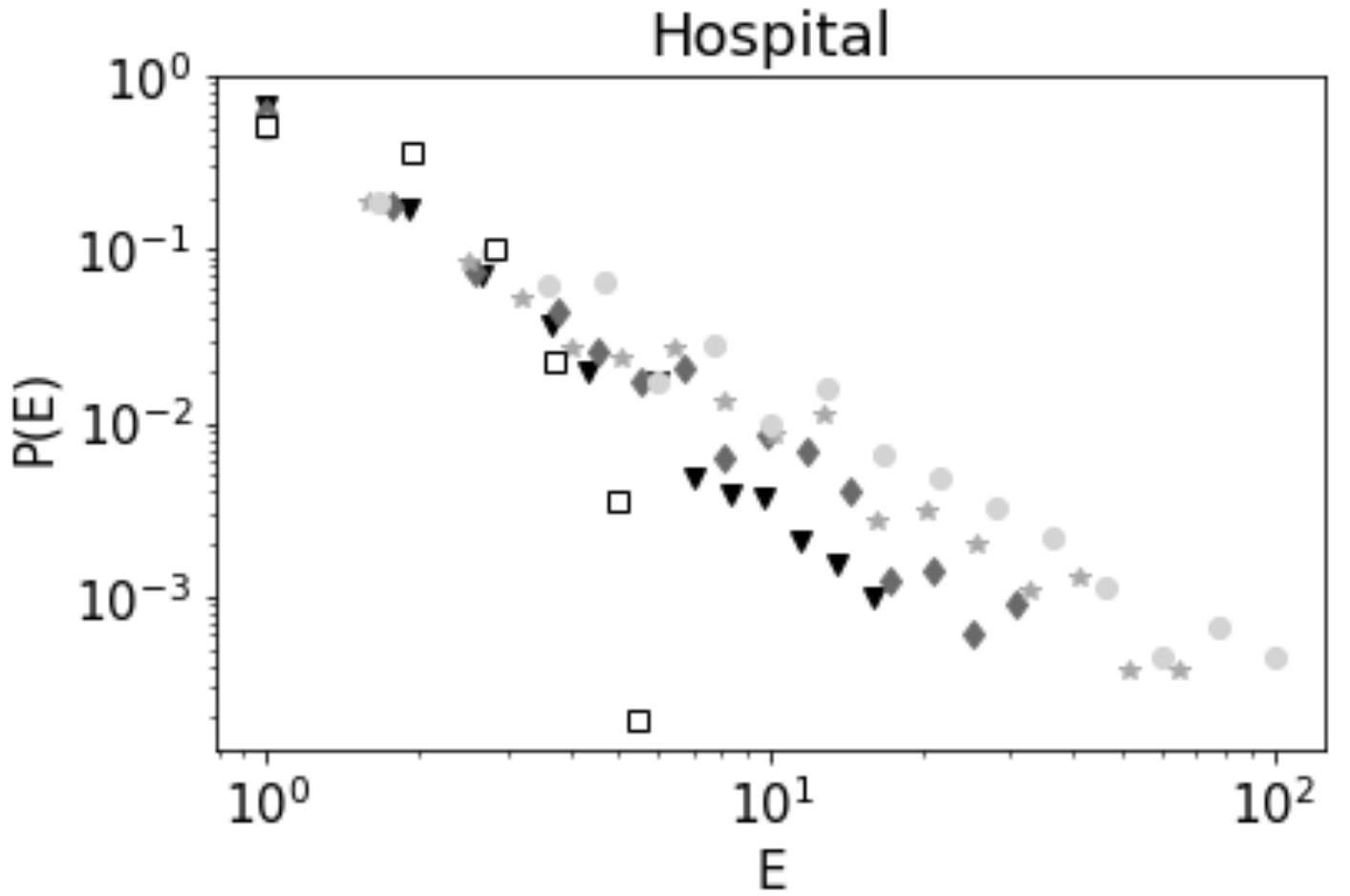}
\hspace{-1mm}
\includegraphics[width=0.32\columnwidth]{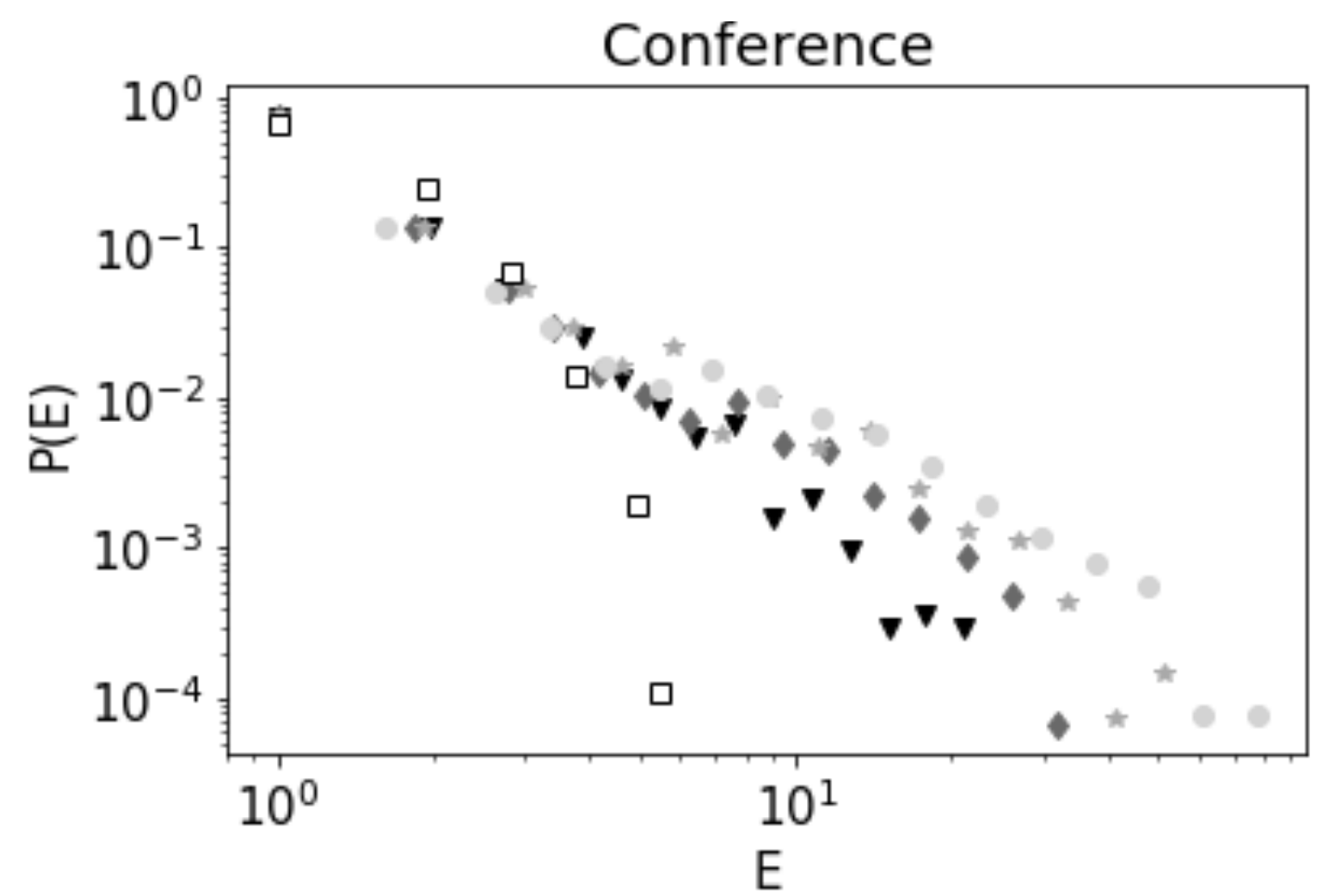}
\caption{The three panels report the number of events' distribution $P(E)$ in the three social settings, analogously to Fig.~3 in main text, with the difference that here the events are counted considering traditionally pairwise interactions. In details, simultaneous interactions involving a closed group of people are not counted as higher-order interactions but are separated into multiple pairwise interactions. This procedure, which ignores the higher-order nature of social interactions, corresponds to the standard method of analysis related to graphs instead of hypergraphs.
For all the three datasets we observe a clear bursty behavior denoted by the power-law shape of the distributions. The distributions are quite independent on choice of the temporal gap $\Delta t$ and differ from the distribution obtained by the null model (empty symbols), which shows a clear exponential behavior.
The null model is obtained with the procedure described in Section Methods of the main text, with $\Delta t=120$ seconds.
}
\end{figure*}

\newpage

\vspace*{2cm}
\centering
\subsection*{Duration time of events}
\vspace{2cm}

\begin{figure*}[ht]
\centering
\hspace*{-10mm}
\includegraphics[width=1.1\columnwidth]{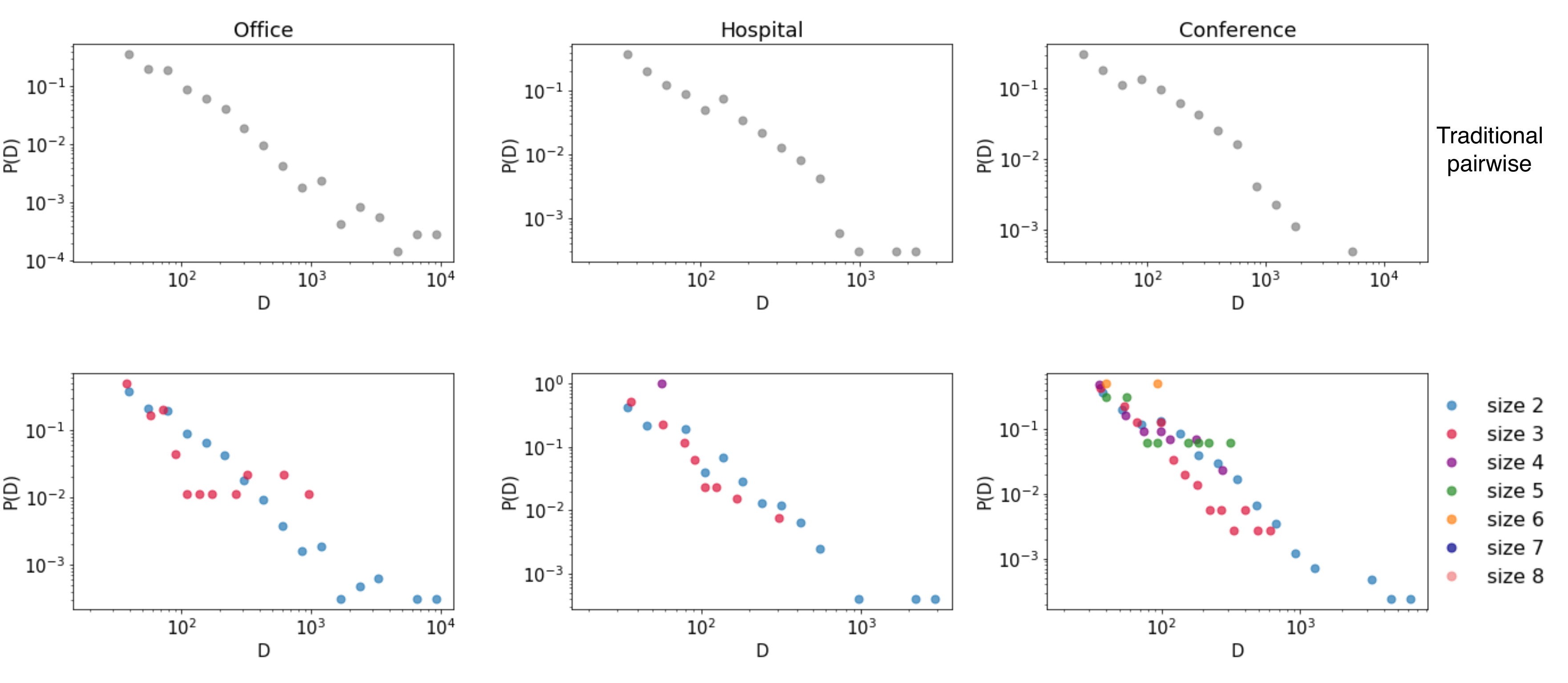}
 \caption{The duration time of events, $D$, in the traditionally pairwise case corresponds to the length in time spanning from the beginning of an interaction between two people to the end of it. In the higher-order framework instead it represents the time between the appearance of a hyperedge and the disappearance of it, whether it disaggregates in a smaller group or it enlarges its size becoming a hyperedge of a different order.
 The duration of a higher-order interaction therefore corresponds to the amount of time during which the hyperedge remains unchanged. During the existence of a hyperedge additional links with external nodes can appear but  if they do not transform it in a hyperedge of a larger size we consider it unchanged.
 The figure is organized in panels where the first line of figures shows the distributions of event durations for the three datasets (here the events are counted as traditionally pairwise interactions and not as hyperedges). Analogous figures are also reported in \cite{starnini2017robust}.
 The second line shows instead how the same distributions appear in a higher-order framework, where interactions are separated according to their size and reported with different colors.
 We observe that smaller sizes of interactions show duration distributions similar to those found for traditionally pairwise interactions. Larger sizes of interactions are quite difficult to compare with the others due to the lack of statistics. 
 }
 \label{PD}
\end{figure*}

\newpage 

\vspace*{2cm}
\centering
\subsection*{Inter-event times}
\vspace{2cm}

\begin{figure*}[ht]
\centering
\hspace*{-10mm}
\includegraphics[width=1.1\columnwidth]{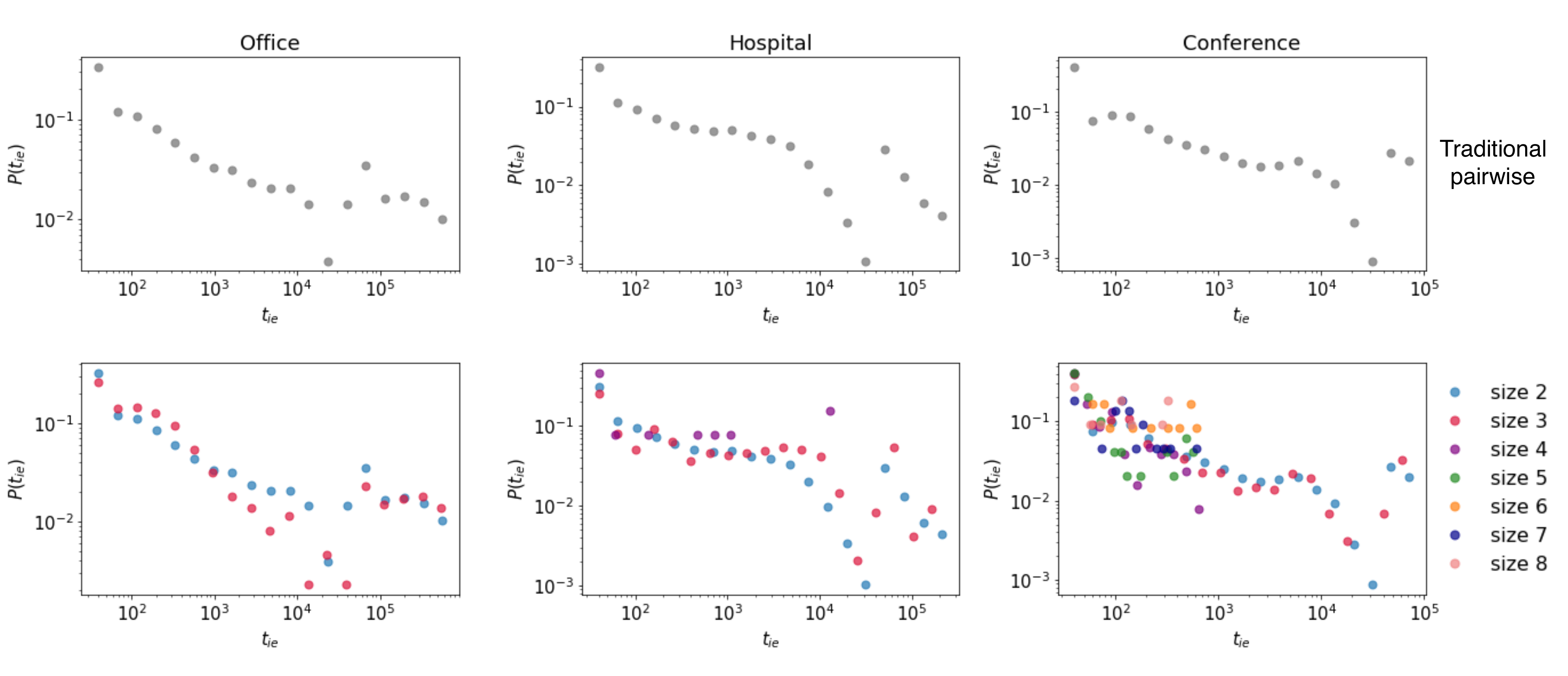}
 \caption{The inter-event time, $t_{ie}$, is defined as the time elapsing between the end of an event and the beginning of a subsequent event (if existing) involving exactly the same people. Both in the traditional and in the higher-order framework, it corresponds to the time between two different consecutive appearances of the same edge or hyperedge.
 The figure is organized in panels where the first line shows the distributions of inter-event time for the three datasets, where the events are counted as traditionally pairwise interactions and not as hyperedges. Analogous figures are also shown in \cite{karsai2012universal}  and \cite{starnini2017robust}.
 The second line shows instead how the same distributions appear in a higher-order framework, where interactions are separated according to their size and reported with different colors.
 We observe that the distributions found for smaller sizes of interactions are similar to those found for traditionally pairwise interactions. Larger sizes of interactions are quite difficult to compare with the others due to the lack of statistics. 
 }
 \label{Ptie}
\end{figure*}

\newpage

\vspace*{2cm}
\centering
\subsection*{Mean size of interactions vs popularity}
\vspace{2cm}

\begin{figure*}[ht]
\centering
\hspace{-4mm}\includegraphics[width=0.34\columnwidth]{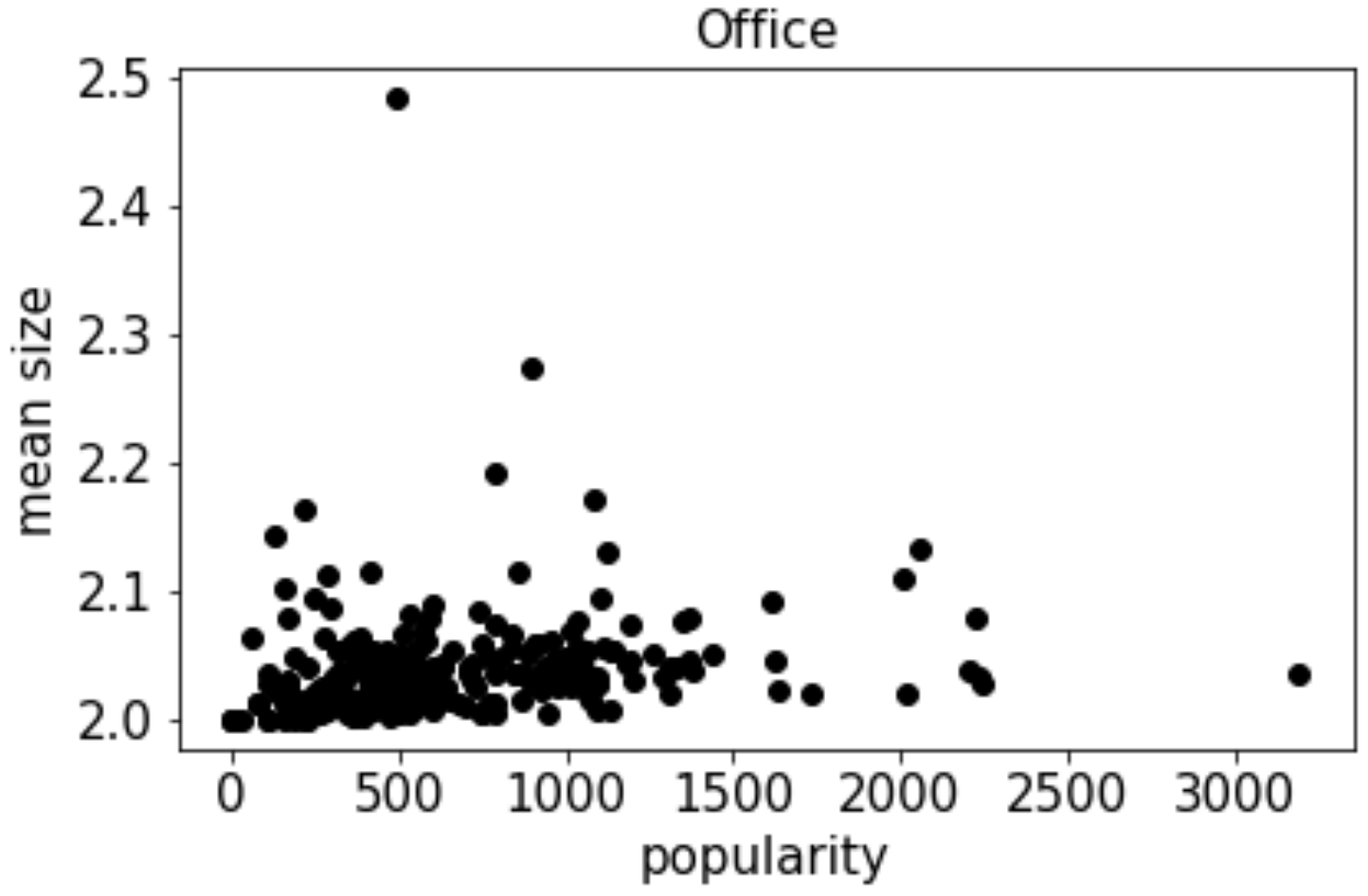}
\hspace{-2mm}
\includegraphics[width=0.34\columnwidth]{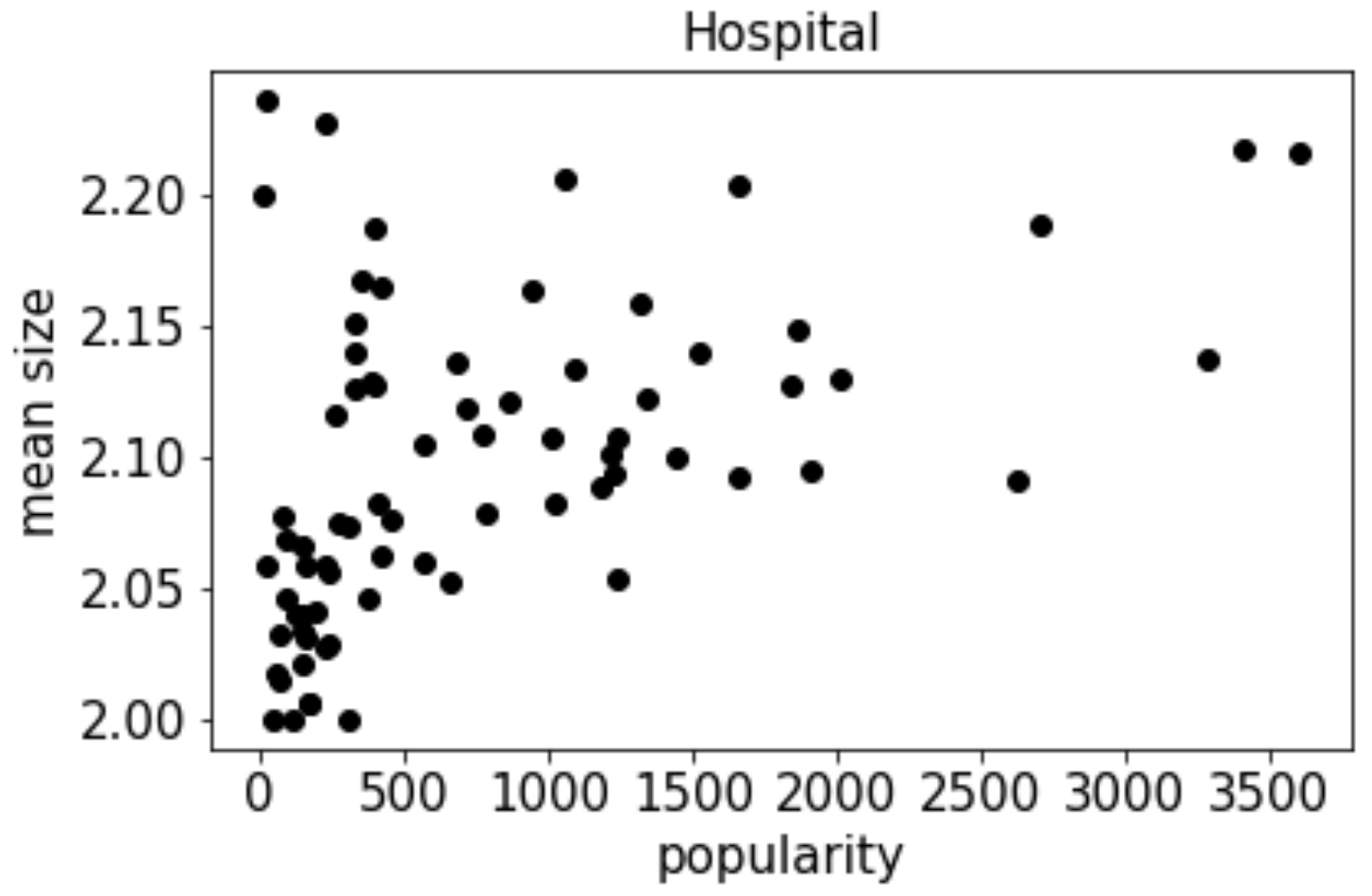}
\hspace{-2mm}
\includegraphics[width=0.34\columnwidth]{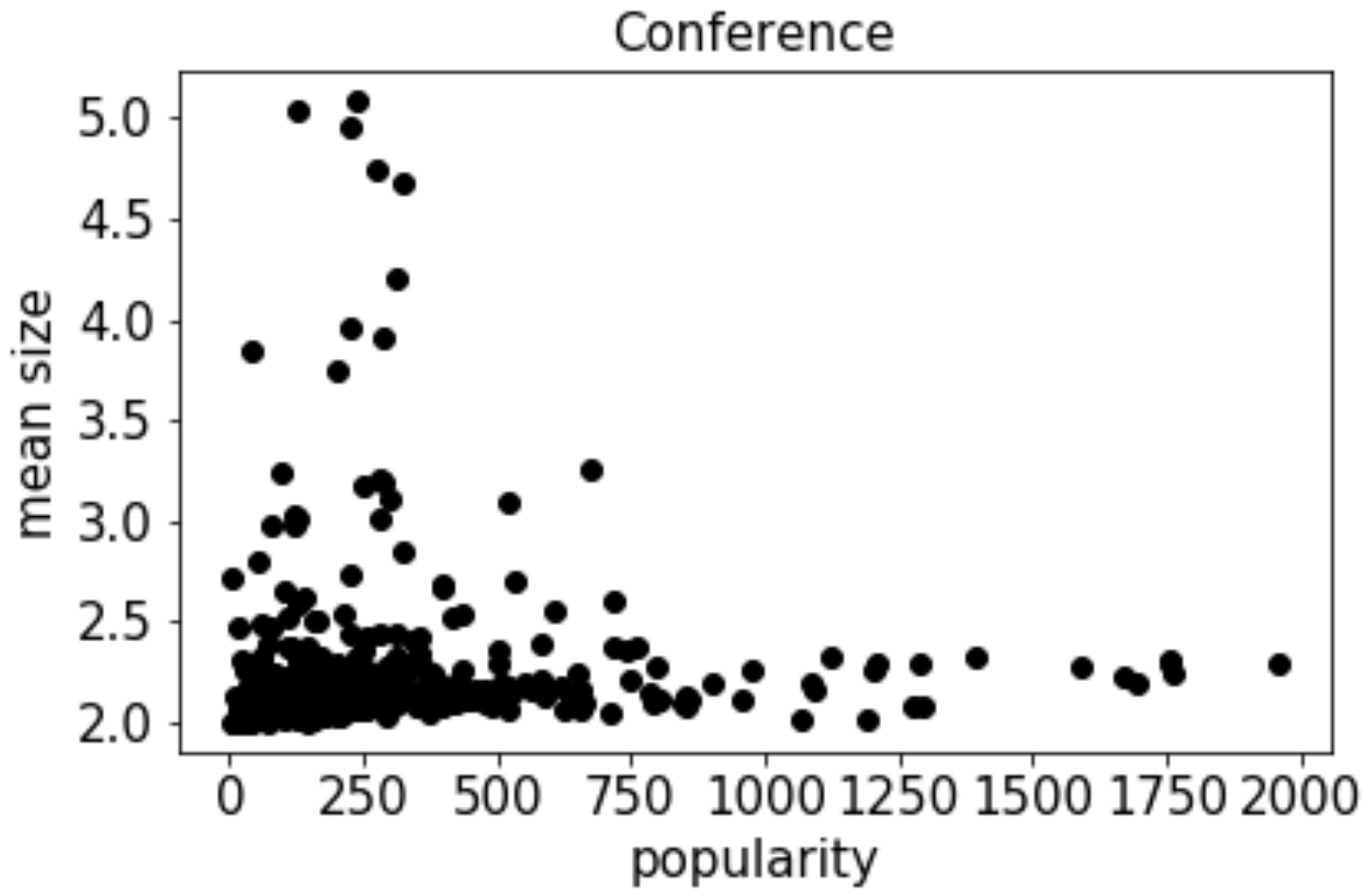}\\
 \caption{The three panels show, for the three datasets, the  relation between popularity of each node in the temporal hypergraph and the average size of its relations. The popularity of a node is defined as the total number of  interaction events where the node is involved, independently oF their size.
 Again, we observe similar behaviors in the Office and the Hospital dataset, where the most popular nodes tend to have larger interactions. Instead, in the Conference setting the largest hyperedges are found for nodes with few interaction events: these people tend to have only few but big interactions, maybe people just coming to the conference to give or attending a big talk. Finally, people with a larger amount of interactions tend to participate to small groups, probably people discussing with many others but not participating to big events.
 }
 \label{fig_pop}
\end{figure*}

\end{document}